\documentclass[%
reprint,
superscriptaddress,
showpacs,preprintnumbers,
bibnotes,
amsmath,amssymb,
aps,
pra,
floatfix
]{revtex4-1}

\usepackage{graphicx}
\usepackage{dcolumn}
\usepackage{bm}
\usepackage{braket}
\usepackage[utf8]{inputenc}
\usepackage[percent]{overpic}
\usepackage{xcolor}

\begin{document}


\title{Probing the out-of-equilibrium dynamics of two interacting atoms}

\author{Tim Keller}
\affiliation{Theoretische Physik, Universität des Saarlandes, D-66123 Saarbrücken, Germany}

\author{Thomás Fogarty}
\affiliation{Theoretische Physik, Universität des Saarlandes, D-66123 Saarbrücken, Germany}
\affiliation{Quantum Systems Unit, Okinawa Institute of
Science and Technology Graduate University, Onna-son, Okinawa 904-0495, Japan}



\date{\today}

\begin{abstract}
We study the out-of-equilibrium dynamics of two
interacting atoms in a one-dimensional harmonic trap after a quench
by a tightly pinned impurity atom. We make
use of an approximate variational calculation called the Lagrange-mesh
method to solve the Schrödinger equation as a function of inter-particle interaction and impurity quench strength. We investigate the out-of-equilibrium dynamics
by calculating the Loschmidt echo which quantifies the irreversibility of the system following the quench, while its probability distribution after long times can be used to identify distinct dynamical regimes. These quantities are related to the spectral function which describes the full dynamical spectrum, and we show through a thorough examination of the parameter space the existence of distinct scattering states and collective oscillations. This work demonstrates how these dynamics are strongly dependent
on the interaction strength between the atoms and may be tuned to observe the establishment of the orthogonality catastrophe in few-body systems.
\end{abstract}

\pacs{67.85.-d, 03.75.Kk, 03.65.Yz}
\maketitle


\section{Introduction}
Advances in optical trapping of cold atoms have allowed for unprecedented manipulation over the size of these quantum systems such that the number of atoms being trapped can be exactly controlled \cite{kinoshita,Jochim2}. Experiments can now explore these few-body systems where the interactions can be tuned via Feshbach resonances to create strongly correlated states \cite{jochim,Jochim3}. Exciting out-of-equilibrium dynamics can lead to distinct oscillations of the wavefunction which are dependent on the interaction modified energy structure of these systems \cite{atomicinterferometer,beating1,beating2,beating3,quenching,Zinner,zollner1,zollner2,misty,misty2}. Probing these oscillations has been proposed and implemented recently \cite{Qubit_probe,mossyoverlap,Goold1,Goold2,Demler,Demler2,Demler3,QubitLL, QubitFS, QubitTH} by using impurity qubits that can trigger these dynamics and which are strongly correlated with the many-body systems being measured. The out-of-equilibrium dynamics of these systems is imprinted on the rate of decoherence of the qubit and can be extracted through interferometric measurements. One can use this scheme to extract information about the system as the full excitation spectrum is obtained. 

In this paper we investigate using an impurity to probe the dynamics of two particles confined to a one dimensional harmonic trap. The particles interact via a contact interaction which leads to non-trivial shifts in the energy levels of the atoms \cite{analyticalsolution} resulting in complex dynamics after a sudden quench of the impurity coupling strength \cite{quenching, Qtherm}. Using numerical tools we quantify this dynamics by calculating a survival probability known as the Loschmidt echo (LE) and its spectral components \cite{zanardi,unitaryequilibration}. Its applications range from understanding decoherence \cite{LE1,cecilia,cecilia2,bruno} and quantum phase transitions \cite{QPTLE3,LE2,QPTLE,QPTLE2,Goold3}, as an indicator of the orthogonality catastrophe (OC) \cite{mossyoverlap,DeDom} and in the study of non-equilibrium quantum thermodynamics \cite{Qthermo,Qtherm,Goold1}. Experimentally, the LE is a measurable quantity in NMR setups \cite{NMR1,NMR2} and can be extracted through Ramsey interferometry \cite{Ramsey,JENS,mossyoverlap,Goold1}. We characterise distinct dynamical classes of the two particle system which are dependent on its interaction and the impurity coupling strength, and we show that depending on the sign of the impurity interaction the system can exhibit distinct scattering and bound state dynamics which can be inferred from the spectral function. 

The paper is organised as follows: In Section \ref{prelim} we introduce the system and the method by which we evaluate the Hamiltonian and calculate the out-of-equilibrium dynamics. In Section \ref{echo} and Section \ref{spectral} we discuss the results of the numerical calculations of the LE and the spectral function. Finally in Section \ref{Conclusion} we conclude our findings and in the Appendix we outline the numerical techniques which we used in this paper.

\section{Model Hamiltonian}
\label{prelim}
We consider a system of two identical bosonic atoms which are trapped in a harmonic potential. Due to strong trap frequencies in two perpendicular directions the atoms are restricted to motion only along the axial direction and can be regarded as one-dimensional. The initial Hamiltonian $\mathcal{H}_{i}$ of the system before the quench reads 
\begin{equation}
\mathcal{H}_{i}=\sum_{j=1}^{2}\left( -\frac{\hbar^{2}}{2m}\frac{\partial^{2}}{\partial \chi^{2}_{j}}+\frac{1}{2}m\omega_T^{2}\chi_{j}^{2}\right)+V_{\text{int}}\left(\chi_{1},\chi_{2}\right),
\label{eq:unperturbedham}
\end{equation}
where $\chi_{j}$ is the coordinate of particle $j$, their mass is labeled $m$ and $\omega_T$ denotes the axial trap frequency.
At low temperatures, the boson-boson interaction consists mainly of \textbf{s}-wave scattering and thus the interaction potential $V_{\text{int}}$ can be approximated by a $\delta$-function potential of strength $g_{\text{1D}}$
\begin{equation}
V_{\text{int}}\left(\chi_{1},\chi_{2}\right)\approx g_{\text{1D}}\delta\left(|\chi_{1}-\chi_{2}|\right)\, .
\end{equation}
The coupling constant $g_{\text{1D}}$ is a tunable parameter which can be modified by exploiting Feshbach resonances of the scattering length $a_{3D}$ or by changing the transverse confinement $d_{\perp}=\sqrt{\hbar/m\omega_{\perp}}$ and has the following form \cite{olshanii}
\begin{equation}
g_{1D}=\frac{4\hbar^{2}a_{3D}}{md_{\perp}^{2}}\frac{1}{1-C\frac{a_{3D}}{d_{\perp}}} \, ,
\end{equation}
where $\omega_{\perp}$ is the trap frequency in the perpendicular directions and $C=\zeta(\frac{1}{2})\approx 1.4603$ is a constant \cite{jochim}.

\begin{figure}
\includegraphics[width=\columnwidth]{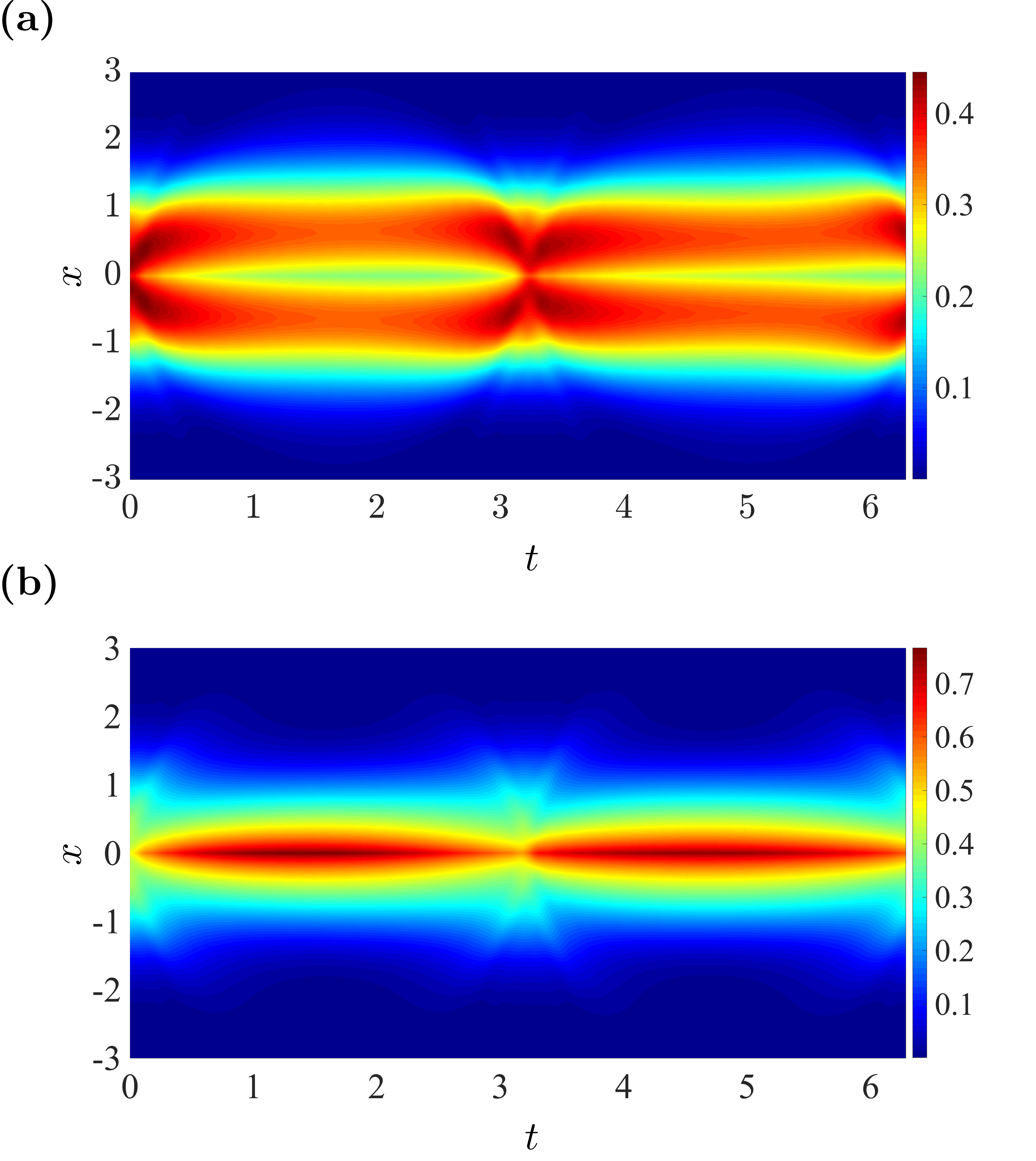}
\caption{Evolution of the single-particle densities for an interaction of $g=2.5$ and a quenched impurity coupling of \textbf{(a)} $\kappa=0.7$ and \textbf{(b)} $\kappa=-0.7$.}
\label{fig:densities}
\end{figure}

We will study the out-of-equilibrium dynamics following an instantaneous quench of the two particle groundstate with an impurity
\begin{equation}
\mathcal{H}_f=\mathcal{H}_i+\Theta(\tau) V_{\text{imp}} \;,
\end{equation}
where $\Theta(\tau)$ is the Heaviside step function and $V_{\text{imp}}$ is a potential which describes the interaction with the impurity that has the form of a $\delta$-function potential barrier of height $\kappa_{0}$ that is located centrally in the trap
\begin{equation}
V_{\text{imp}}\left(\chi_{1},\chi_{2}\right)=\sum_{j=1}^{2}\kappa_{0}\delta(\chi_{j})\, .
\label{Vimp}
\end{equation}
We assume that the impurity is a tightly pinned atom which has two distinct internal levels, $|0\rangle$ and $|1\rangle$, such that it describes a qubit. For simplicity, we assume that the qubit ground state $|0\rangle$ does not interact with the bipartite system, while the excited state $|1\rangle$ interacts with the coupling strength $\kappa_0$, which can be controlled with Feshbach resonances. Therefore by exciting the qubit to its excited state we can suddenly trigger its interaction with the bipartite state and create out-of-equilibrium dynamics following this quench.

In the following, we scale all lengths by $a=\sqrt{\hbar/\left(m\omega_T\right)}$, all energies by $\hbar\omega_T$ and give the time $\tau$ in units of the inverse trapping frequency $\omega_T^{-1}$. Thus the scaled dimensionless quantities are given by $x_{j}=\chi_{j}/a$, $\kappa=\kappa_0/(a\hbar\omega_T)$, $g=g_{\text{1D}}/(a\hbar\omega_T)$ and $t=\tau\omega_T$ which leads to the following scaled Hamiltonians

\begin{equation}
\begin{split}
\tilde{\mathcal{H}}_{i} =\sum_{j=1}^{2} & \left(  -\frac{1}{2}\frac{\partial^{2}}{\partial x _{j}^{2}}+\frac{1}{2}x_{j}^{2} \right) + g\delta (|x_{1}-x_{2}|)\, ,
\end{split}
\label{eq:quenchedham1}
\end{equation}

\begin{equation}
\begin{split}
\tilde{\mathcal{H}}_{f} =\sum_{j=1}^{2} & \left(  -\frac{1}{2}\frac{\partial^{2}}{\partial x _{j}^{2}}+\frac{1}{2}x_{j}^{2} + \kappa\delta (x_{j})\right)\\ & + g\delta (|x_{1}-x_{2}|)\, .
\end{split}
\label{eq:quenchedham2}
\end{equation}

The time independent Schr\"{o}dinger equation  $\tilde{\mathcal{H}}_{i} \psi_n(x_1,x_2)=E_n\psi_n(x_1,x_2)$ can be solved analytically by introducing center-of-mass and relative coordinates \cite{analyticalsolution}, however the quenched system's Hamiltonian $\tilde{\mathcal{H}}_{f}\phi_n(x_1,x_2)=E_n^{'}\phi_n(x_1,x_2)$ lacks such a treatment and we therefore solve it numerically. We accomplish this by using the Lagrange-mesh method which is an approximate variational calculation \cite{lmm} and is further explained in the Appendix. 

We study the dynamics of the system after the quench by expressing the time dependent wavefunction in terms of the eigenstates of $\tilde{\mathcal{H}}_f$ such that
\begin{equation}
\Psi(x_{1},x_{2},t)=\sum_{n=0}^{\infty}a_{n}\phi_{n}(x_{1},x_{2})e^{-iE'_{n}t},
\label{Psit}
\end{equation}
where
\begin{equation}
a_{n}=\int_{-\infty}^{+\infty}\int_{-\infty}^{+\infty}\phi_{n}^{*}(x_{1},x_{2})\psi_{0}(x_{1},x_{2})dx_{1}dx_{2}\, ,
\label{eq:an}
\end{equation}
is the overlap of the final Hamiltonian's eigenstates with the initial ground state $\psi_0(x_1,x_2)$. The dynamics of the system can be qualitatively understood through the time evolution of the single particle density
\begin{equation}
\rho(x,t)=\int_{-\infty}^{+\infty}\left|\Psi(x,x_{2},t)\right|^{2}dx_{2}\, .
\end{equation}
We consider the excited state of the impurity to have either a repulsive (positive $\kappa$) or an attractive (negative $\kappa$) coupling to the bipartite state, which will result in different dynamics after the quench. In Fig. \ref{fig:densities} the evolution of the single-particle density is shown for an interaction of $g=2.5$ and different impurities of strength $\kappa=\pm 0.7$. For a repulsive impurity coupling the quench imparts kinetic energy to the particles causing the density to spread out to the trap edges with a pronounced dip appearing at the position of the impurity. The density follows a quasi-harmonic motion as the particles interaction and the impurity potential significantly alters the energy structure of the state. Nonetheless, a partial revival of the density is visible on short time scales occurring around $t\approx\pi/\omega_T$. For attractive quenches, the density is localised in the center of the trap as the state is attracted to the site of the impurity and forms a bound state which causes high frequency oscillations in the single particle density. When the strength of the quench is small the dynamics caused by the static impurity quench show comparable characteristics to that of a mobile impurity considered in \cite{quenching}, as the effect of the impurity motion on the bipartite state is negligible in this case. However, for large quenches the two models begin to diverge as the motion of the impurity is enhanced by the quench of the strong interspecies interaction. These impurity density fluctuations will then be written on the bipartite density and may be observed as high frequency oscillations.

\section{Loschmidt echo}
\label{echo}
We calculate the LE to investigate the complex dynamics which stem from the sudden impurity quench, and to understand its dependence on the quench coupling strength and the interactions between the two particles. The LE describes the reversibility of a given dynamical evolution, whereby it illustrates the disparity between two states as a result of imperfect time reversal, and it is therefore closely related to the fidelity \cite{fidelity}. The LE can be measured experimentally through Ramsey interferometry as explained in \cite{mossyoverlap}, whereby a $\pi/2$ pulse is applied to the impurity qubit such that the overall system becomes a correlated state of the form $(|0\rangle \otimes e^{-i \tilde{\mathcal{H}}_i t}|\psi_0\rangle+|1\rangle \otimes e^{-i \tilde{\mathcal{H}}_f t}|\psi_0\rangle)/\sqrt{2}$. This triggers out-of-equilibrium dynamics in the bipartite state as it is quenched by the sudden coupling to the excited state of the qubit. A measurement of the probability of the qubits state allows one to extract $\nu(t)$ which describes its decoherence and is related to the LE, $\mathcal{L}(t)$, through
\begin{equation}
\mathcal{L}(t)=|\nu(t)|^2=|\braket{\psi_{0}|e^{i\tilde{\mathcal{H}}_{\text{f}}t}e^{-i\tilde{\mathcal{H}}_{\text{i}}t}|\psi_{0}}|^{2}\, ,
\end{equation}
therefore one can extract the dynamics of the bipartite system through a measurement of the qubit. The LE is essentially the time dependent overlap of the initial state evolving with and without the effects of the impurity quench, and can be rewritten using Eq.~\eqref{Psit} in the compact form
\begin{equation}
\mathcal{L}(t)=\left|\sum_{n}\left|a_{n}\right|^{2}e^{i(E_0-E'_{n})t}\right|^{2}\, .
\label{eq:le_our_case}
\end{equation}
This shows that the evolution of the LE is determined by the difference in energy between the initial state $E_0$ and the excitation spectrum of the quenched Hamiltonian $E'_n$, and is weighted by the factors $a_n$. At $t=0$ the LE is unity as the two states are equivalent, while for $t>0$ the quench disturbs the density of the bipartite state resulting in out-of-equilibrium dynamics and a decay of the LE, with strong quenches resulting in a temporal state which is far from the initial state and the possibility of reaching a dynamical orthogonal state signaled by $\mathcal{L}(t)\rightarrow0$ \cite{mossyoverlap,quenching,DeDom}. Due to finite size effects from the harmonic trapping potential, revivals of the state will be observed as the density refocuses in the center of the trap causing an increase of the LE. The frequency of these revivals depends on the energy level shifts $E_0-E'_{n}$ and can give a indication of the energy structure of the system: evenly spaced energy levels $E'_n$ will mean sharp revivals and have $\mathcal{L}(t)\rightarrow 1$ at the revival times, while irregularly spaced levels will dephase and have broad and diminished revivals. 

\begin{figure}
\includegraphics[width=\columnwidth]{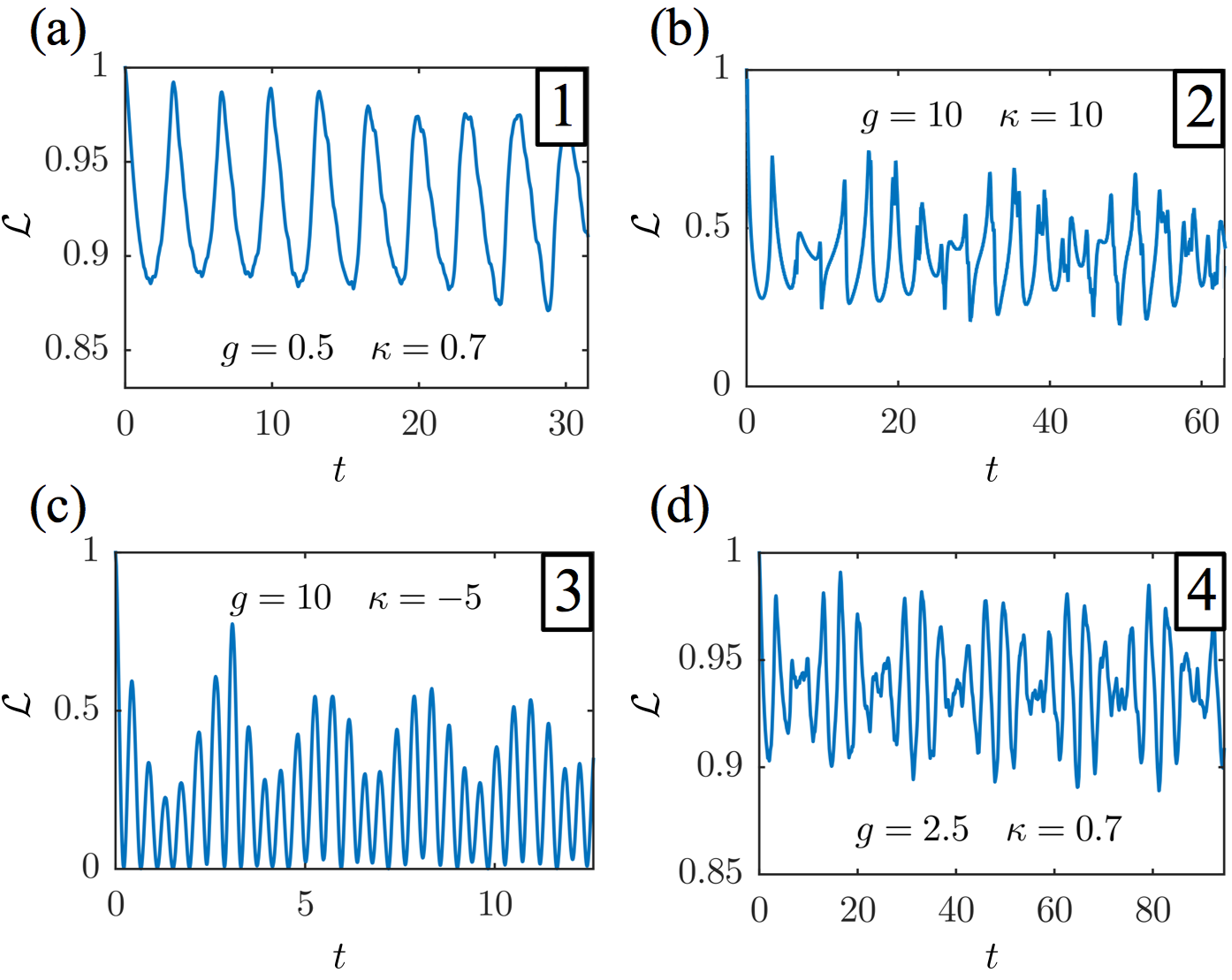}
\caption{Evolution of the LE following the impurity quench for different values of the bipartite interaction $g$ and the impurity coupling $\kappa$. The four panels each show a distinct behaviour depending on the choice of parameters and possess oscillations on different time scales.}
\label{fig:snippets}
\end{figure}

For infinitely repulsive interactions the system is in the Tonks-Girardeau (TG) limit and may be solved using the Fermi-Bose mapping theory\cite{fermibosemapping,girardeauwrightmomdis}, allowing us to treat the infinitely repulsive bosonic two particle wavefunction as a system of two non-interacting fermions. There are known solutions to the problem of a single particle in a $\delta$-function split trap which allow us to solve this system exactly \cite{analyticalsolution,murphy}. Due to the form of the mapping, the LE is identical for both the TG and the fermionic systems. In the case of the latter the LE can be written in terms of the single particle overlaps $A_{mn}(t)=\langle \varphi'_n(x,t) \vert \varphi_m(x,t)\rangle$, such that
\begin{equation}
\mathcal{L}(t)=\left| A_{00}(t)A_{11}(t) - A_{01}(t)A_{10}(t) \right|^2
\label{EchoTG}
\end{equation}
where $\varphi(x,t)$ [$\varphi'(x,t)$] are the harmonic oscillator [quenched] time-dependent single-particle states. As we only take the two lowest single particle states into account, the only state influenced by the impurity is $\varphi'_0$, while the wavefunction of the first excited state is zero at the position of the impurity and is hence unaffected by it, meaning that $\varphi_1=\varphi'_1$. Therefore $A_{11}(t)=1$, and due to the orthonormality of the set of eigenstates of the harmonic oscillator the last term in Eq.\eqref{EchoTG} vanishes, which means we need to only consider the contribution of the evolution of the groundstate fermion to the LE in this case. The LE for two TG particles under the influence of the impurity is given by 
\begin{equation}
\mathcal{L}(t)=\left|A_{00}(t)\right|^{2}=\left|\sum_{n}\left|\tilde{a}_{n}\right|^{2}e^{i(\epsilon_0-\epsilon'_{n})t}\right|^{2}\, ,
\end{equation} 
where $\epsilon_0$ [$\epsilon'_n$] are the single particle energies of the harmonic oscillator [quenched] states and $\tilde{a}_{n}$ denotes the time-independent overlaps $\braket{\varphi'_{n}(x)|\varphi_{0}(x)}$. For larger TG systems under the influence of the impurity described by Eq.~\ref{Vimp} we need to only consider the contribution of the even states to the LE as the odd states are unaffected. However, for finite sized impurities both even and odd states will be altered by the impurity and will therefore need to be taken into account when calculating the LE.


\begin{figure}
\includegraphics[width=\columnwidth]{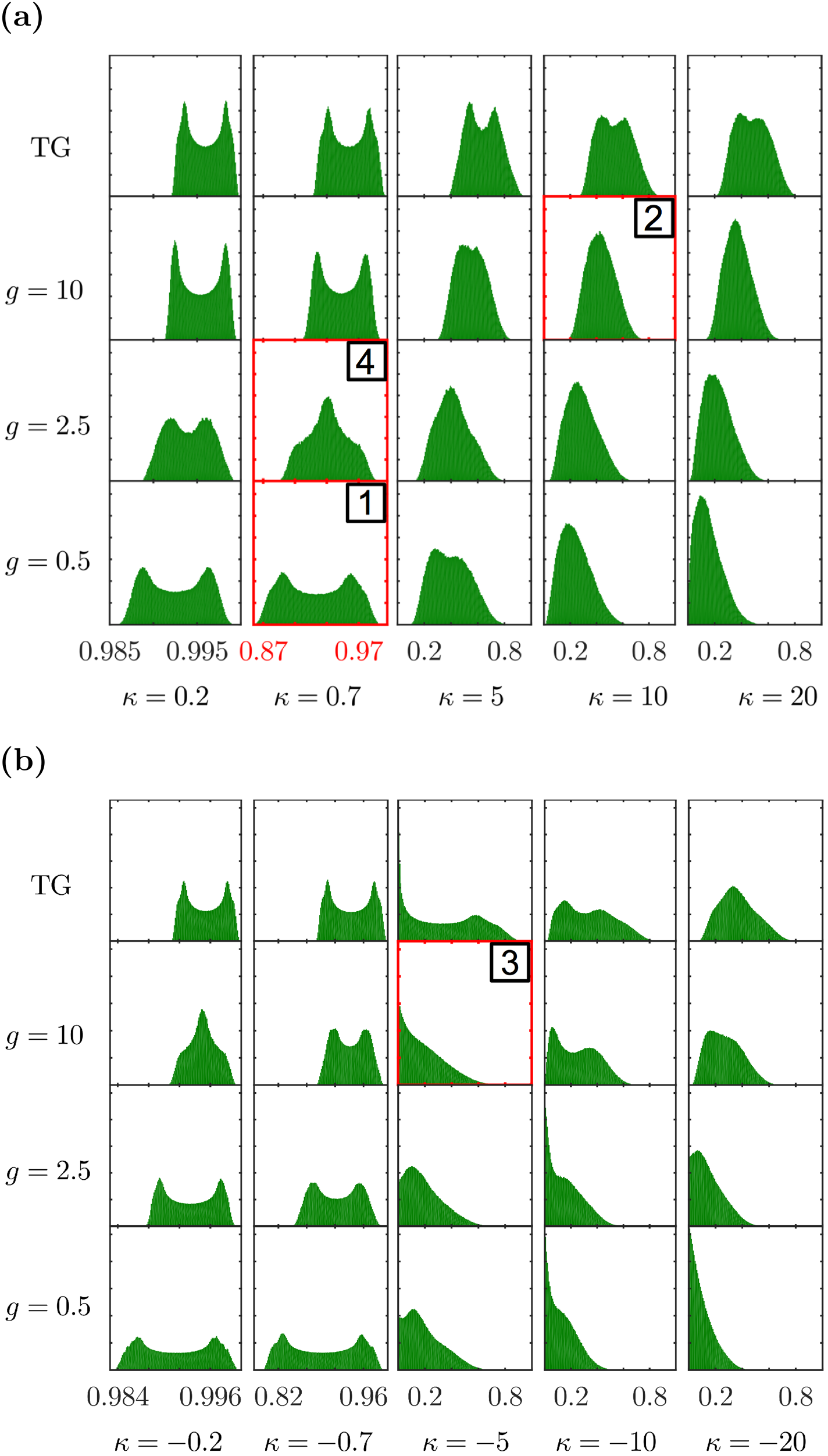}
\caption{Probability distributions of the LE following the quench for different values of the bipartite interaction strength $g$, and for the quench coupling strength (a) $\kappa>0$ and (b) $\kappa<0$. The LE for the numbered distributions are plotted in Fig.~\ref{fig:snippets}.}
\label{fig:distributions}
\end{figure}
In general, the LE exhibits an oscillatory behaviour for weak quenches (Fig. \ref{fig:snippets}(a)) similar to that shown in \cite{quenching} for the mobile impurity. For strong repulsive impurity quenches the LE is proportionately smaller in magnitude and its evolution is more complex as it now involves higher energy components which destroys the periodicity of the revivals (Fig. \ref{fig:snippets}(b)). In the previously discussed situation of the mobile impurity, a high frequency oscillation would also be present in the evolution of the LE as the impurity density fluctuates between the two strongly interacting particles, whereas in the situation of the static impurity this is not seen. For strong attractive impurity quenches (Fig. \ref{fig:snippets}(c)), the LE shows high frequency oscillations with rapidly changing amplitudes and periodically achieves orthogonality ($\mathcal{L}=0$). Finally, a distinct beating pattern can be observed in the LE for specific parameter combinations, which we have only observed for small interactions (Fig. \ref{fig:snippets}(d)). The specific response of the LE in these figures will be explained in detail in the following sections.

\subsection{Probability distribution of the LE }
While it is clear from Fig.~\ref{fig:snippets} that the evolution of the LE acquires some characteristic shape and behaviour on short times, it is difficult to say so quantitatively. Therefore, we numerically evaluate the probability distribution of the LE
\begin{equation}
P(y)=\lim_{T\rightarrow \infty}\frac{1}{T} \int_{0}^{T} \delta(\mathcal{L}(t)-y) dt \;,
\end{equation}
for those $y$ which coincide with the values of the LE. This distribution is calculated after evolving the system for a long period of time such that all the intricacies of the evolution is captured (typically on time scales of $t\approx 2\pi \times 1600 \omega_T^{-1}$). Experimentally this could be achieved by creating many replicas of the initial state and evolving them subjected to same quenched Hamiltonian, whereby measurements on the individual states at different times would build the probability distribution \cite{unitaryequilibration}. The distribution of the LE is shown in Fig.~\ref{fig:distributions} for different parameter combinations ranging from weak to strong regimes of interaction and impurity quench strength, as well as the analytically solvable case of the TG pair. The distributions of the four plots of the LE from Fig.~\ref{fig:snippets} are highlighted by red borders and their corresponding numbers in Fig. \ref{fig:distributions}. Each of these different quenches have distinct distributions depending on the chosen parameters and we can approximately separate them into four categories: 
\begin{enumerate}
\item a \textbf{double-peaked distribution} when the LE oscillates quasi-periodically after a small quench
\item a \textbf{Gaussian distribution} when the LE possesses a complex noisy shape for strong quenches
\item an \textbf{exponential distribution} for strong quenches which create dynamical orthogonality 
\item and a \textbf{winged distribution} when there is a beating pattern in the LE for a small quench.
\end{enumerate}
The other parameter combinations show slight deviations or mixtures of these basic distributions as the system transitions smoothly from one regime to the next, however a clear trend is noticeable as a function of the quench strength which we will now discuss. 

\begin{figure}
\includegraphics[width=0.9\columnwidth]{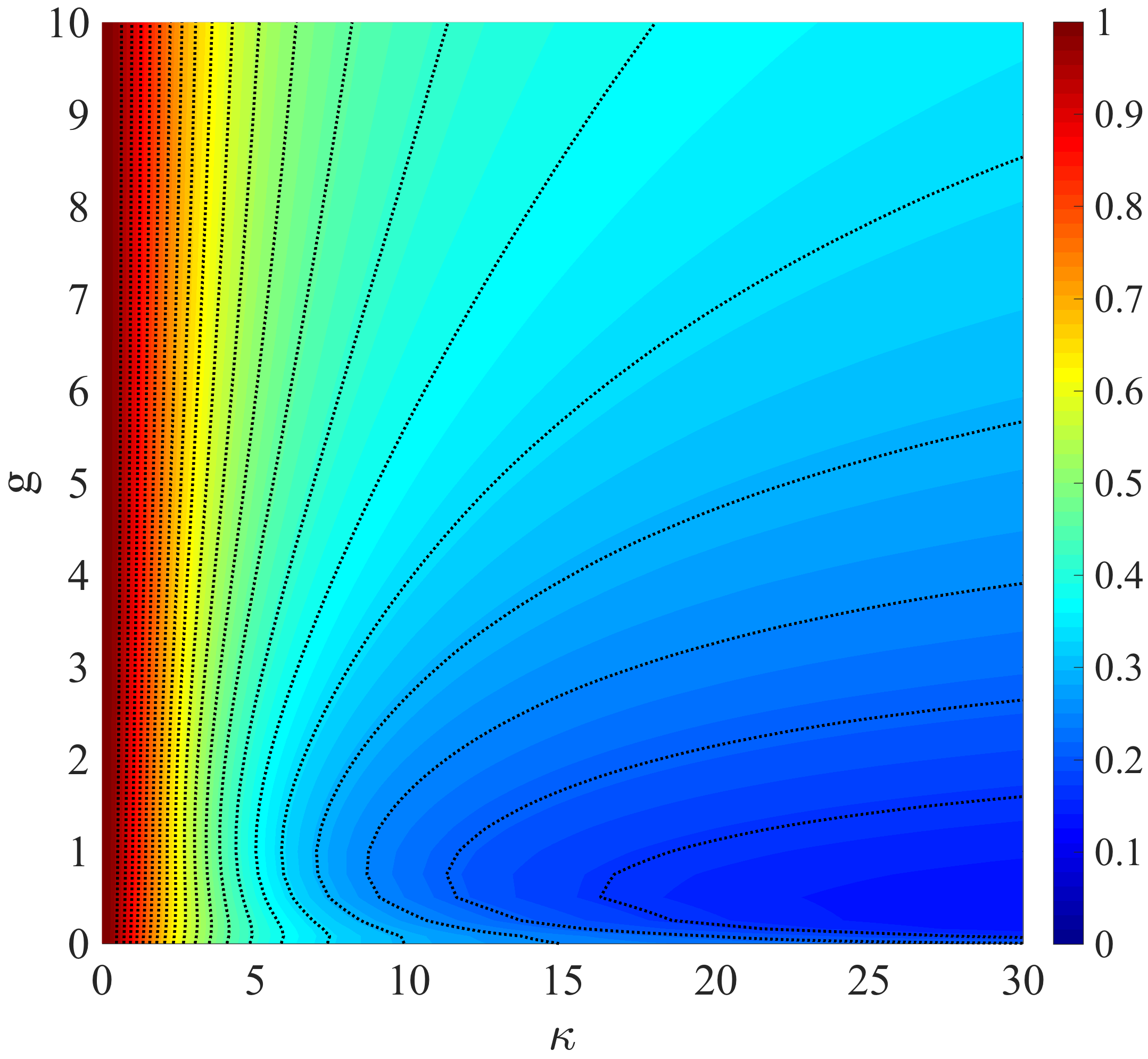}
\caption{Mean of the LE, $\overline{\mathcal{L}}$, as a function of impurity coupling strength $\kappa$ and the bipartite interaction strength $g$, where the contour lines are added as a guide to the eye.}
\label{AvLE}
\end{figure}

In general, the LE will decrease with increasing $\kappa$ due to the growing destructive influence of the quench which reduces the overlap between the initial and quenched states. This is visible in the changing of the scales with $\kappa$ in Fig.~\ref{fig:distributions} and the behaviour of the average LE, $\overline{\mathcal{L}}=\sum_n|a_n|^4$, plotted in Fig.~\ref{AvLE}. It is clear that the response of the system to the impurity is dependent on the interaction strength and will effect the values the LE can take. The shift of the LE distribution to higher values for growing interaction (for constant $\kappa$) is a result of the effect of the point-like interactions on the two particle wavefunction. For $g$ large, the diagonal $x_1=x_2$ in $\psi(x_1,x_2)$ diminishes as strong repulsive interactions force the particles apart. This in turn lowers the density in the center of the trap where the impurity is situated, thus diminishing the impact of the impurity quench on the initial state. In the extreme case of the TG pair, the quench only affects the center-of-mass component of the wavefunction as the relative component is zero at the origin. This reduces the impact of the quench significantly on the bipartite state resulting in a larger mean value of the LE and a smaller width of the probability distribution. Small interactions are needed to achieve lower LE and to approach a dynamical OC, whereby the relative and center-of-mass motion are coupled due to the presence of the impurity, and the density of the bipartite state at the impurity is only slightly reduced. The destructive effect of the impurity quench is the most pronounced for $g\approx 1$ in Fig.~\ref{AvLE} where the minimum of the mean LE is found. Due to the presence of the revivals in the LE this temporal averaging will limit it magnitude to a finite value larger than zero, however between these revivals the LE is seen to vanish showing that weak interactions can be used to approach the OC in few-body systems \cite{quenching}.

\begin{figure*}
\begin{minipage}{0.49\textwidth}
\begin{flushleft}
\textbf{(a)}
\end{flushleft}
\includegraphics[width=\columnwidth]{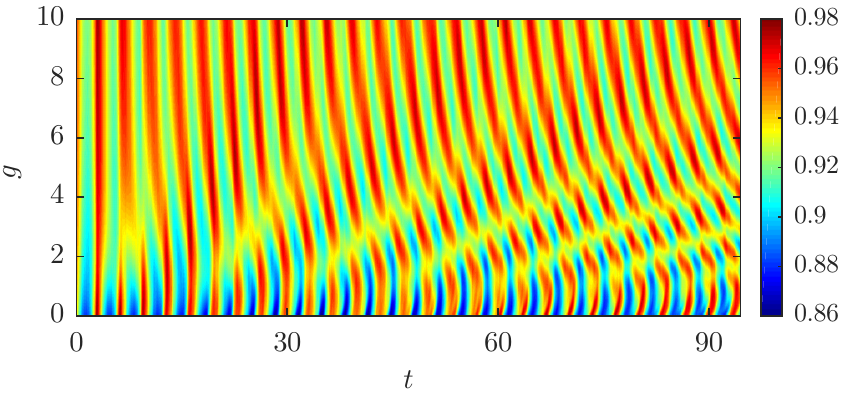}
\end{minipage}
\hfill
\begin{minipage}{0.49\textwidth}
\begin{flushleft}
\textbf{(b)}
\end{flushleft}
\includegraphics[width=\columnwidth]{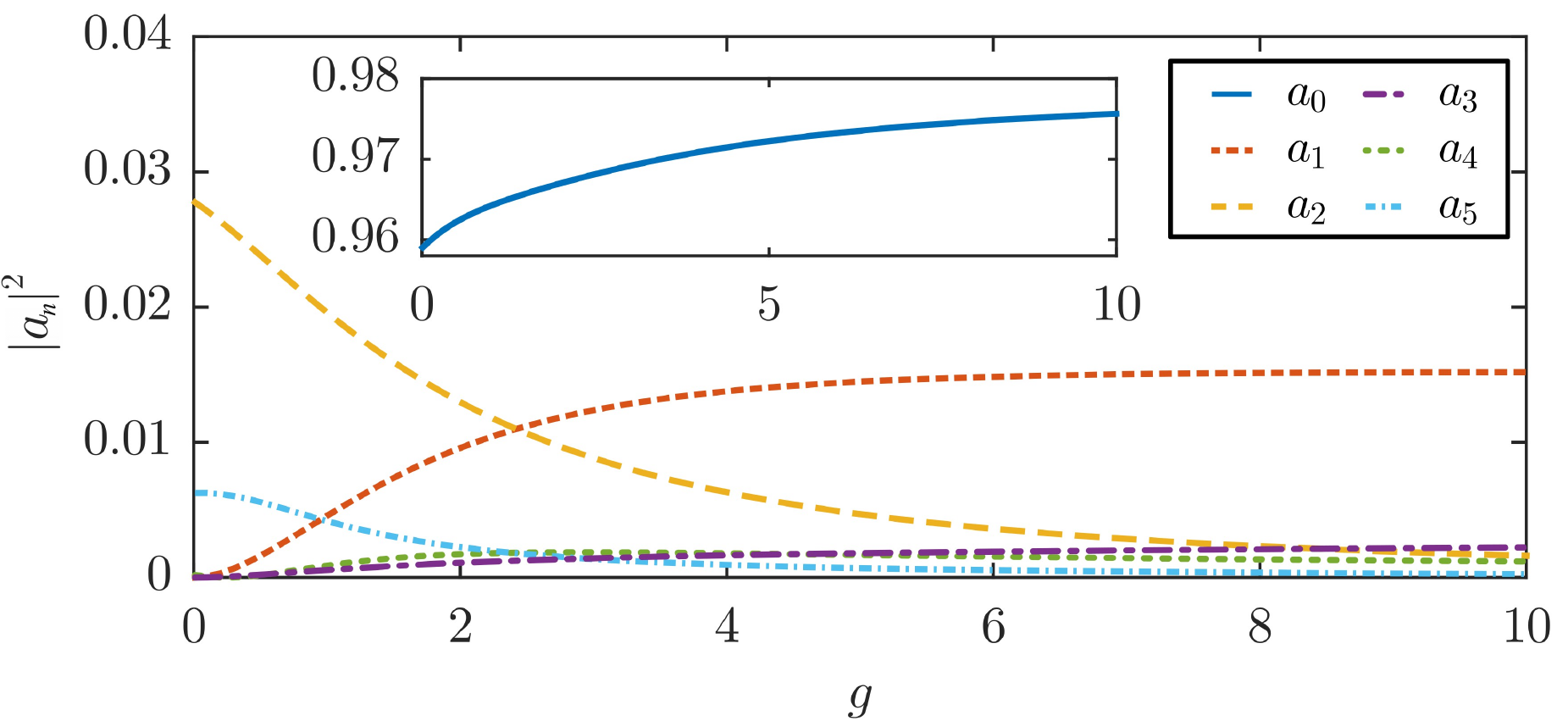}
\end{minipage}
\caption{\textbf{(a)} Evolution of the LE as a function of the bipartite interaction $g$ after an impurity quench of coupling strength $\kappa=0.7$. The frequency of the oscillations in the LE is dependent on $g$, and for $g\approx2.5$ fringes appear in the evolution highlighting the beating pattern observed in Fig.~\ref{fig:snippets}(d). \textbf{(b)} Evolution coefficients $|a_n|^2$ (calculated using Eq. \eqref{eq:an}) as a function of $g$. The LE beatings occur around the crossing between the coefficients $a_{1}$ and $a_{2}$.}
\label{fig:revg}
\end{figure*}

In the following we will discuss in more detail the four specific distributions.

\subsubsection{Double-peaked distribution}
As shown in Fig.~\ref{fig:densities}, the quench imparts kinetic energy to the particles which excites periodic oscillations of the state, whereby it expands and then contracts to the initial density at the trap center. As the bipartite state is kicked to the trap edges it is at its least dense, such that the overlap with the initial state will be minimal and thus have a low value of the LE. Whereas when the state refocuses at the trap center the LE will be maximal as its density increases and will exhibit a sharp revival (Fig.~\ref{fig:snippets}(a)). The probability distribution of the LE will therefore be bunched at these two points resulting in a double-peaked distribution which also gives us important information about the frequency of these oscillations. As the quench strength is small there is only a low number of excitations and the energy structure of the quenched state is close to that of the harmonic oscillator, therefore these two peaks in the distribution can be resolved \cite{unitaryequilibration}.

\subsubsection{Gaussian distribution}
If the quench is strong there will be a large number of excitations with energies that are far from the harmonic spectrum. This means that these excitations will dephase as the state evolves after the quench and the revivals of the LE will be diminished (see Fig.~\ref{fig:snippets}(b)). This will result in the double-peaked probability distribution becoming blurred into a shape resembling a Gaussian. 

\subsubsection{Exponential distribution}
As discussed previously, by tuning the interaction strength in the bipartite state the signature of OC may be observed for strong quench strengths. In this case the LE will be vanishingly small and will be punctuated with diminished revivals of finite magnitude, therefore the corresponding probability distribution of the LE will have a large peak at $\mathcal{L}=0$ with an exponentially decaying tail. This effect can be seen to emerge for a positive quench of $\kappa=20$ at $g=0.5$ and for multiple values of the negative quench in Fig.\ref{fig:distributions}, such as $g=10$ and $\kappa=-5$ which is shown in Fig.~\ref{fig:snippets}(c).

\subsubsection{Winged distribution}
\label{winged}
Finally, we will discuss the winged distribution which only exists for a reduced parameter range of small quenches, and can be identified as a central peak in the LE probability distribution surrounded by two lobes. This unique distribution describes a regular beating pattern visible in the evolution of the LE for $g=2.5$ and $\kappa=0.7$ (see Fig.\ref{fig:snippets}(d)), where the system oscillates with a breathing mode. The beating in the dynamics is a direct result of the small finite interactions which causes nontrivial energy level shifts away from the harmonic spectrum. This can result in periodic temporal evolution involving two distinct frequencies \cite{beating1, beating2, beating3}. The emergence of this beating is shown in Fig.~\ref{fig:revg}~\textbf{(a)} where the evolution of the LE is plotted as a function of $g$ for a small quench in $\kappa$. For $g\approx0$ the revivals of the LE are periodic and regular, as the initial energy spectrum is close to harmonic and the small quench magnitude introduces only a minor anharmonicity to the system. For $g>0$ the formation of beatings can be seen in the appearance of fringes in the revivals of the LE. This beating is the result of a resonance between two distinct frequencies which become equally dominant in the evolution operator of the state. The weights of these frequencies, $|a_n|^2$, are plotted in Fig.~\ref{fig:revg}~\textbf{(b)} for the six largest contributions to the evolution and is dominated by the energy shift of the ground state, $E'_0-E_0$. The beating is the most pronounced, when the coefficients $a_{1}$ and $a_{2}$ are of equal magnitude around $g\approx 2.5$. As $E'_1$ and $E'_2$ are close in energy the envelope of the LE's beating can be visible on long timescales of the order of ten trap periods. In contrast, other resonances of these coefficients, for example $a_1$ and $a_5$, have a much larger energy difference which results in small fluctuations at a much higher frequency, which is not easily visible in the LE. For large interactions ($g\gtrsim 10$) any beating in the LE vanishes due to the system returning to an initial state which possesses a regularly spaced energy spectrum close to that of the TG gas.


\begin{figure*}
\includegraphics[width=2\columnwidth]{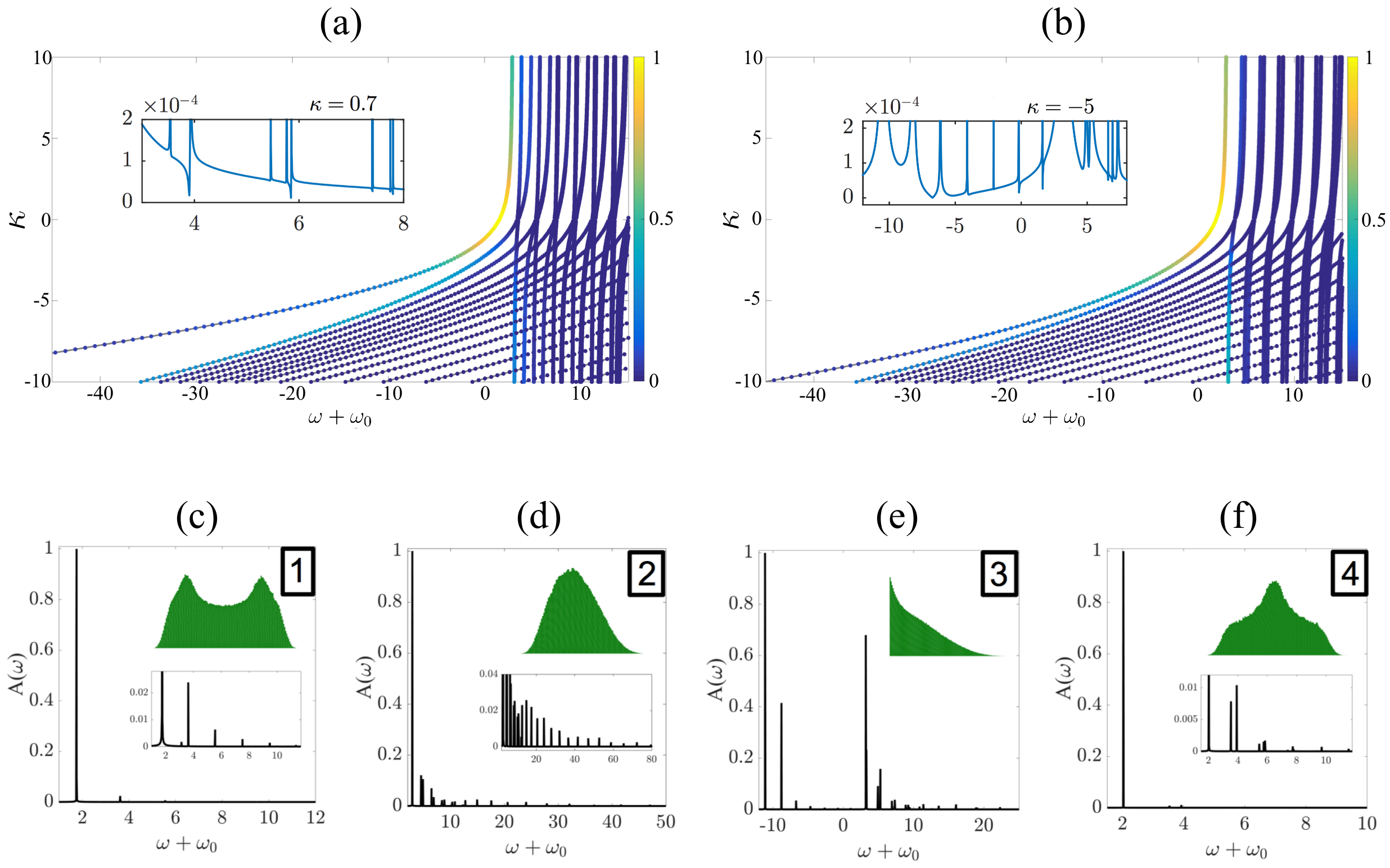}
\caption{Spectral function for varying $\kappa$ and fixed interactions of \textbf{(a)} $g=2.5$ and \textbf{(b)} $g=10$. The frequency $\omega$ is in units of $\omega_T$ and the spectra have been scaled by the height of the ground state spectral peak $E'_0-E_0$ for $\kappa\approx0$. The color scale of the points represent the height of the spectral peaks. The insets show close-ups of the spectral functions for the given parameters, whereby Fano resonances are visible at the base of the peaks. \textbf{(c-f)} Spectral functions for the four basic cases displayed in Fig. \ref{fig:snippets} and their resulting LE probability distributions. The frequency axes have been shifted by $\omega_{0}$ which is the frequency of the unperturbed ground state for the respective interaction.}
\label{fig:spectra}
\end{figure*}


\section{Spectral function}
\label{spectral}
The spectral function is a powerful tool which can be used to understand the out-of-equilibrium dynamics following the quench, whereby it describes the statistics of the initial state and the subsequent quantum dynamics by providing the excitation spectrum of the system. In its discrete form it is given by
\begin{equation}
A(\omega)=2\pi\sum_{n}\left|a_{n}\right|^{2}\delta\left(\omega-\omega_{0}+\omega'_{n}\right)\, ,
\label{discreteA}
\end{equation}
where $\omega'_{n}$ are the eigenfrequencies of $\tilde{\mathcal{H}}_f$ and $\omega_{0}$ is the groundstate frequency of the initial state. It is related to the work probability distribution describing the work done on/by the system following a quench, therefore its direct measurement can be used to determine quantum thermodynamical statistics \cite{Qtherm, Qthermo}. We numerically evaluate the spectral function by calculating the Fourier transform of the time-dependent overlap of the initial and quenched state $\nu(t)$ \cite{mahan}, such that
\begin{equation}
A\left(\omega\right)=2\pi \: \text{Re} \int_{-\infty}^{\infty}e^{i\omega t}\nu(t)dt.
\label{eq:specfunc}
\end{equation}
Therefore the spectral function is the frequency representation of the LE and should complement its description of the out-of-equilibrium evolution. When calculating Eq. \eqref{eq:specfunc} we simulate suitably large time scales such that we capture all the relevant dynamics and we check that it agrees with the discrete expression in Eq.\eqref{discreteA}.

Fig. \ref{fig:spectra} shows the spectral function for \textbf{(a)} $g=2.5$ and \textbf{(b)} $g=10$. For a repulsive impurity coupling ($\kappa>0$) the excitation frequencies are all positive in nature and are seen to be dominated by the lowest frequency excitation, which is the difference in energy between the initial and final groundstates, $E'_0-E_0$. For increasing quench strength the number of excitations visible in the spectral function is seen to increase as higher energy states are excited by the impurity and begin to play a role in the dynamics. These excitations are comprised of groups of nearly degenerate states which can be generally categorised into separate effective center-of-mass (COM) and relative (REL) oscillations of the two-body wavefunction. The interaction between the particles only affects the REL states thereby introducing a cusp in the wavefunction at the center of the trap, this will naturally alter the effect of the quench on these states compared to the COM states. For low interactions ($g=2.5$), strong quenches cause the excitations to visibly split into pairs with diverging frequencies as the COM and REL states are effected differently by the impurity. For large interactions ($g=10$) the even states become doubly degenerate with the odd states, meaning the effects of the quench in the strong interaction limit are essentially equal for the COM and REL states, resulting in a minor splitting of the excitation frequencies.  

\begin{figure}
\includegraphics[width=0.9\columnwidth]{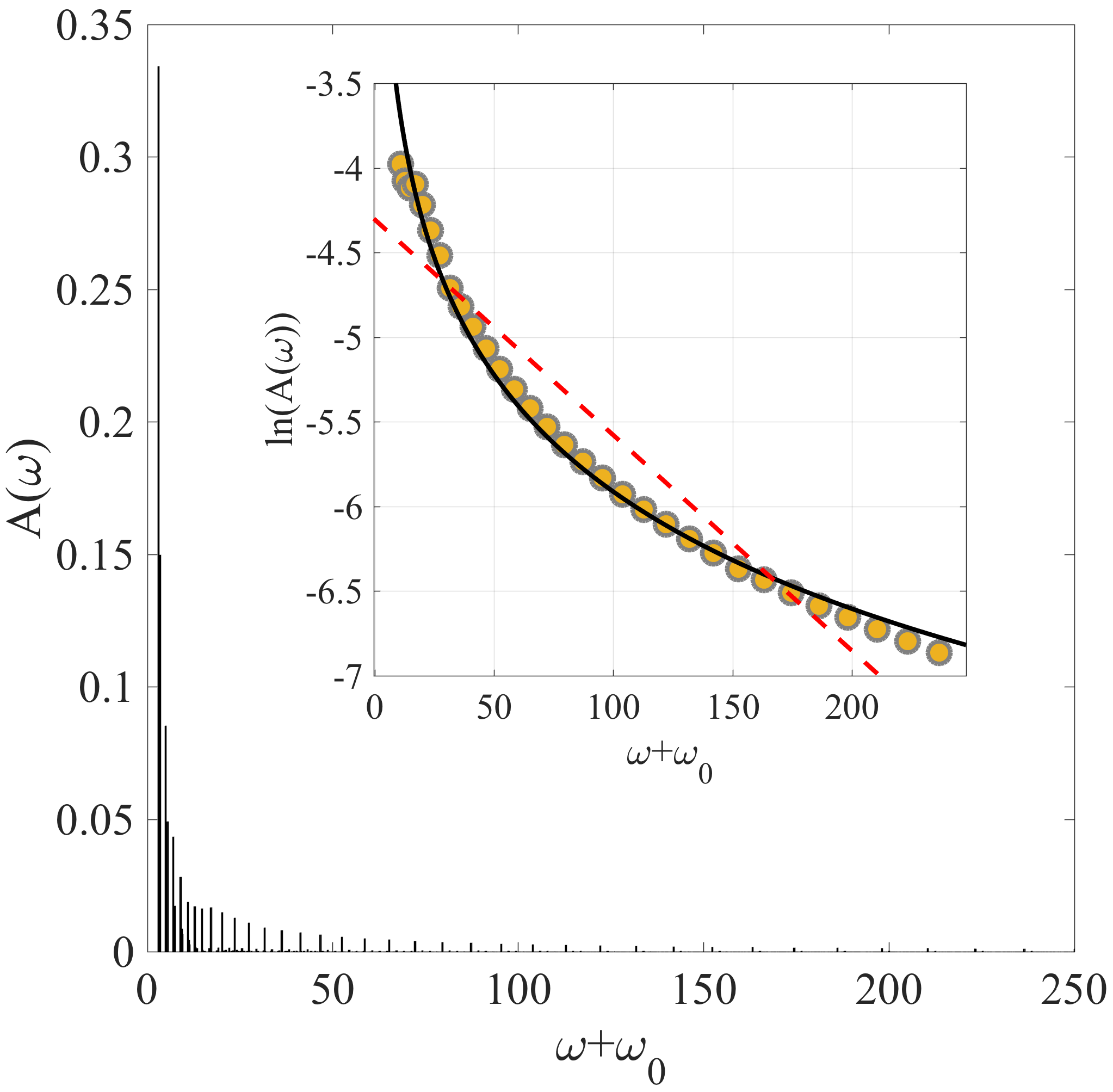}
\caption{Spectral function for $\kappa=30$ and $g=1$ which exhibits the emergence of the OC. The inset shows the logarithm of the spectral tails (dots) with a power law fit (solid line) and an exponential fit (dashed line). The frequency $\omega$ is in units of $\omega_T$.}
\label{fig:OC_spec}
\end{figure}

Similar effects are visible for $\kappa<0$ and $\omega+\omega_0>0$ where the attractive impurity will create positive energy excitations in the system. Owing to the attractive nature of the impurity a more complex dynamics can be witnessed due to the creation of a bound state. This is visualised in the spectrum as the appearance of a second branch of peaks when $\omega+\omega_0<0$ in Fig.\ref{fig:spectra} (a-b) \cite{Demler3}. These negative frequency excitations describe the motion of the bound state which is tightly confined at the position of the impurity, and its energy can be seen to decrease unbounded as $\kappa$ is decreased. Therefore the bipartite state subject to an attractive impurity will display two distinct dynamics: (i) trap dominant positive energy excitations which oscillate in the harmonic potential with a time scale on the order of the inverse trap frequency $1/\omega_T$ and are not bound to the impurity, (ii) bound state dominant excitations that have mainly negative energy which oscillate in the bound state and whose time scale is determined by the energy of the bound state. There is a cross-over region where both sets of excitations exist and we observe that it is dependent on the interaction in the bipartite state. For weak interactions ($g=2.5$) the coupling of the COM and REL motion means it is favourable to form a bound state at the impurity. There is a large contribution of this bound state to the dynamics as the dominant peaks of the spectral function have negative energy. For strong interactions ($g=10$) the repulsion between the particles causes a density dip at the point of the impurity reducing the ability to form a bound state, therefore the crossover region in this case has mainly trap dominant excitations. The insets in panel Fig.~\ref{fig:spectra} (a) and (b) show the spectral function at a fixed barrier height for the respective interaction. The upper part of the spectra have been cut off to better visualise the base of the excitation peaks. These insets reveal, that most peaks exhibit Fano resonances, a feature which is typical in the spectrum of two coupled oscillators \cite{fano1}.

Fig.~\ref{fig:spectra} \textbf{(c-f)} shows the spectral functions of the four characteristic cases with the respective LE's shown in Fig.~\ref{fig:snippets}. The double-peak, exponential and Gaussian distributions show spectra that are in good agreement with those reported in \cite{unitaryequilibration} which reflect these distribution shapes. In panel (e) the exponential distribution is a result of the splitting of the spectral function into two separate branches. For larger attractive impurity strengths the two branches become more and more separate and the resulting LE probability distribution is a superposition of the distributions of these individual branches. This also explains the many different LE probability distributions occurring in the strong impurity coupling regime, as the individual dynamics discussed in the previous paragraph will merge. In panel (f) the beating responsible for the winged probability distribution is caused by the spectral peaks of approximately the same height around $\omega+\omega_{0}\approx 3.5$ as explained previously in Sec.~\ref{winged}.

Finally, to explore the onset of the OC which was observed in Fig~.\ref{AvLE}, we calculate the spectral function for $\kappa=30$ and $g=1$ (see Fig~.\ref{fig:OC_spec}). The spectral function possesses a prominent peak at low frequency followed by rapidly diminishing peaks in the high frequency tails. Indeed, the birth of the OC manifests in the power law decay of these spectral tails which are plotted in the inset of Fig.~\ref{fig:OC_spec}. The tails are fitted to a power law decay (solid line) which shows good agreement, especially in comparison with the exponential decay (dashed line) which would be observed for a system at equilibrium \cite{mahan,mossyoverlap}.

\section{Conclusion}
\label{Conclusion}
We have investigated the effect of interparticle interactions on the out-of-equilibrium dynamics of two atoms following a sudden quench from the coupling with an impurity. A thorough analysis of this dynamics was carried out through numerical calculations of the LE, which describes the survival probability of the initial state and contains information about the excitations in the system. Finite interactions were shown to result in nontrivial dynamics of the bipartite state which can be characterised into four distinct probability distributions of the LE and were found to have a direct role in observing the emergence of the OC in few-body systems. A breathing mode was also observed in this system for weak interactions which is the result of an interference between two non-equilibrium excitations. To fully understand the complexity of this system the spectral function was calculated, this illustrated two different evolution dynamics when the impurity quench was attractive: trap and bound-state oscillations. 

We foresee our results to be relevant to the study of few- and many-body systems as it allows complex out-of-equilibrium dynamics to be categorised and understood in the statistics of the LE. Recent experiments where neutral impurities interact with a Bose gas of a different atomic species show promising setups for our work, as the use of species-selective potentials and Fesbach resonances allow for precise control of the impurity position and tailoring of its interaction \cite{widera,scelle,catani}. Our analysis is also readily applicable to recent experiments with impurities in Fermi gases, whereby Ramsey interferometry has been performed on $^{40}$K impurities following an interaction quench with a $^6$Li Fermi sea \cite{Demler}, and in \cite{Jochim2} few-body systems of $^6$Li atoms which are deterministically created in a one-dimensional trap and probed with an impurity atom. And a recent demonstration of the ability to manipulate single $^{87}$Rb atoms into ordered arrays \cite{Lukin} showcases the ability to deterministically create and control these small system sizes discussed in our work.

\begin{acknowledgements}
We would like to thank Giovanna Morigi and Thomas Busch for their support and discussions, and TF thanks John Goold for insightful exchanges. We acknowledge support from the German Research Foundation (DFG, DACH project Quantum crystals of matter and light) and BMBF (Qu.com).
\end{acknowledgements}

\appendix*
\section{Lagrange-mesh method}
\label{app:lmm}
The Lagrange-mesh method is a numerical method similar to \textit{pseudo-spectral} or \textit{Discrete Variable Representation} methods \cite{lmm}. In a first step the examined region $[a,b]$ is divided into $N$ mesh points $x_{1},\ldots,x_{N}$ and a Gauss quadrature for the numeric integration of arbitrary functions $g(x)$ on the mesh has to be determined 
\begin{equation}
\int_{a}^{b}g(x)dx\approx\sum_{k=1}^{N}\lambda_{k}g(x_{k}) \, .
\label{eq:quadratur}
\end{equation}
A basis of $N$ Lagrange-functions has to be chosen, that fulfill the following interpolation and orthogonality conditions
\begin{eqnarray}
f_{i}(x_{j})= \lambda_{i}^{-1/2}\delta_{ij} \quad \forall \; i, j\\
\label{eq:cond1}
\int_{a}^{b} f_{i}^{*}(x)f_{j}(x)dx=\delta_{ij}\, .
\label{eq:cond2}
\end{eqnarray}
Together with the ansatz
\begin{equation}
\Psi (x)=\sum_{i=1}^{N}c_{i}f_{i}(x) \, ,
\label{eq:ansatz1d}
\end{equation}
where $c_{i}=\lambda_{i}^{1/2}\Psi (x_{i})$, this leads to the equations
\begin{equation}
\sum_{j=1}^{N} \left( T_{ij}+V(x_{j})\delta_{ij}\right)c _{j}=Ec_{i}
\label{eq:lmm1d}
\end{equation}
that have to be solved for a general one-dimensional problem with $T_{ij}=\braket{f_{i}|T|f_{j}}$ and $V_{ij}\approx V(x_{i})\delta_{ij}$ is approximated by means of the Gauss quadrature. In this work we used a cartesian mesh with unity spacing of the mesh points at $x_j=j$, $j=-\frac{1}{2}(N-1),\ldots,\frac{1}{2}(N-1)$. The Fourier basis functions read
\begin{equation}
f_{i}(x)=\frac{1}{N}\frac{\text{sin}[\pi(x-x_{i})]}{\text{sin}[\pi(x-x_{i})/N]}
\label{eq:lagrangefunc}
\end{equation}
and $\lambda_{i}=1 \quad \forall  \; i$. In this case the kinetic energy terms are given by
\begin{equation}
T_{ij}=\begin{cases}\frac{\pi^{2}}{6}\left(1-\frac{1}{N^{2}}\right) &i=j \\ 
(-1)^{(i-j)}\frac{\pi^{2}}{N^{2}}\frac{\text{cos}[\pi(i-j)/N]}{\text{sin}^{2}[\pi(i-j)/N]} &i\neq j \, .
\end{cases}
\label{eq:kineticmatrix2}
\end{equation}
Applying this approach to the two-particle case leads to the final equations for our problem
\begin{widetext}
\begin{equation}
\sum_{k,l=1}^{N}\left\lbrace\left(\frac{1}{h^{2}}T_{ik}+V(hx_{k})\delta_{ik}\right)\delta_{jl}+\left(\frac{1}{h^{2}}T_{jl}+V(hx_{l})\delta_{jl}\right)\delta_{ik}+\frac{g}{h}\delta_{il}\delta_{jl}\delta_{kl}\right\rbrace\Psi_{kl}=E\Psi_{ij} \, ,
\label{eq:scaledeqsystem}
\end{equation}
\end{widetext}
where $V\left(x_{i}\right)=\frac{1}{2}x_{i}^{2}+\kappa\delta\left(x_{i}\right)$.
Equation \eqref{eq:scaledeqsystem} is simply the tensor product of two single particle systems with the added interaction term.  Additionally we introduced a scaling parameter $h$ that allows us to adapt the mesh to the region of interest and thus increase the resolution of the method.
\bibliography{draft}
\end{document}